\documentclass{article}
\usepackage{algorithm}
\usepackage{algorithmic}
\usepackage{listings}
\usepackage{amssymb}
\usepackage{amsmath}
\usepackage{amsthm}
\theoremstyle{plain}
\newtheorem{theorem}{Theorem}[section]

\theoremstyle{remark}
\newtheorem{remark}[theorem]{Remark}
\theoremstyle{definition}

\usepackage{appendix}
\usepackage{booktabs}
\usepackage{placeins}
\usepackage{graphicx} 
\usepackage{hyperref}
\providecommand{\keywords}[1]
{
  \small	
  \textbf{\textit{Keywords:}} #1
}
\title{Adaptive Multilevel Splitting: First Application to Rare Event Derivative Pricing }
\author{ Riccardo Gozzo\thanks{PhD Student, Scuola Normale Superiore, Pisa. Work conducted while at the University of Milano-Bicocca.}}
\date{}
\begin{document}
\maketitle
\begin{abstract}
\noindent This work analyzes the computational burden of pricing binary options in rare event settings and introduces an adaptation of the adaptive multilevel splitting (AMS) method for financial derivatives. Standard Monte Carlo is inefficient for deep out-of-the-money binaries due to discontinuous payoffs and low exercise probabilities, requiring very large samples for accurate estimates. An AMS scheme is developed for binary options under Black–Scholes and Heston dynamics, reformulating the rare event problem as a sequence of conditional events. Numerical experiments compare the method to Monte Carlo and to other techniques such as antithetic variables and multilevel Monte Carlo (MLMC) across three contracts: European digital calls, Asian digital calls, up-and-in barrier digital calls. Results show up to a 200-fold computational gain for deep out-of-the-money cases while preserving unbiasedness. To the best of our knowledge, this is the first application of AMS to financial derivative pricing. The approach improves pricing efficiency for rare-event contracts such as parametric insurance and catastrophe-linked securities. An open source Rcpp implementation is provided, supporting multiple discretizations and importance functions.
\end{abstract}

\section{Introduction}
The accurate and efficient pricing of financial derivatives is increasingly critical in modern markets, where advanced numerical methods are required for complex instruments \cite{jrfm12010035}. The computational challenges of rare event simulation extend beyond academic interest, creating bottlenecks that affect market functionality. Inaccurate pricing of low probability events limits the ability of market makers to provide competitive quotes, reducing liquidity for these instruments \cite{Muellerleile_2025}. This difficulty is pronounced in the insurance sector, where parametric products depend on binary triggers linked to observable parameters such as earthquake magnitude or wind speed \cite{undici, LARSSON2023100345}. Computational limitations restrict coverage of catastrophic risks and constrain the development of innovative risk transfer mechanisms in financial and insurance markets. 
\vspace{0.2cm} \\ These challenges are evident in binary options, which share structural similarities with parametric insurance through trigger-based payoffs. Their discontinuous structure pays a fixed amount if the underlying asset crosses a predetermined barrier at expiration and zero otherwise \cite{shreve2004stochastic2, shreve2004stochastic1}. This all-or-nothing feature makes pricing highly sensitive to the probability of rare events, particularly for deep out-of-the-money contracts where accurate tail estimation is critical.
\vspace{0.2cm} \\Addressing these difficulties naturally leads to simulation-based techniques. Monte Carlo methods are widely used for pricing complex derivatives due to their flexibility in high-dimensional settings \cite{glasserman2004monte}. The convergence rate of $O(N^{-1/2})$ creates a computational bottleneck, especially for binary options with low exercise probabilities. Reliable estimation in such cases typically requires millions of paths, rendering crude Monte Carlo impractical \cite{beck2015rareeventsimulation, bucklew2004introduction}.
\vspace{0.2cm} \\ Classical variance reduction techniques attempt to address these challenges. Antithetic variates reduce variance through negative correlation between paired samples \cite{hammersley1956new, glasserman2004monte}, but the theoretical gain is bounded by a factor of two \cite{varredu}. Control variates can be more effective but require auxiliary variables that are both analytically tractable and highly correlated with the target payoff \cite{rasmussen2005control}. For discontinuous payoffs such as binary options, such variables are difficult to construct, limiting applicability.
\vspace{0.2cm} \\ More advanced methods have been developed. Importance sampling modifies the probability measure to increase the frequency of rare outcomes and applies likelihood ratio weighting to remove bias \cite{glasserman2004monte, imps}. Its effectiveness depends on the design of suitable distributions, which is problem-specific and difficult to generalize \cite{swiler2010importance}. Another prominent approach is multilevel Monte Carlo (MLMC), which reduces complexity by combining simulations on coarse and fine discretizations \cite{m1, m2}. While efficient for path-dependent derivatives, MLMC is not tailored to extreme event pricing, focusing instead on reducing overall cost.
\vspace{0.2cm} \\ Recent research combines these techniques to overcome individual limitations. Hybrid methods integrate MLMC with importance sampling to improve efficiency while concentrating sampling in critical regions \cite{BenAlaya17022023, kebaier2018coupling}. Machine learning further enhances importance sampling, with neural networks learning tilting parameters \cite{muller2019neural} and tensor train decompositions enabling high-dimensional distribution approximation \cite{cui2024deep}.
\vspace{0.2cm} \\ This work addresses the computational challenges of binary option pricing by applying the adaptive multilevel splitting (AMS) method \cite{Cérou27022007}. AMS extends classical splitting techniques for rare event simulation \cite{garvels2000splitting} and builds on the foundations of sequential Monte Carlo \cite{doucet2001sequential}. Originally developed in reliability analysis and statistical physics \cite{c1, baars2021application, innes2024adaptive, refId0}, AMS decomposes a rare event into a sequence of more frequent conditional events, transforming a single intractable estimation into multiple tractable subproblems. Although AMS has achieved strong results in other scientific domains, no prior applications are documented in financial derivatives pricing. Recent advances provide theoretical guarantees of unbiasedness and convergence \cite{brehier2016unbiasedness, cerou2016fluctuation, cerou2023adaptive}, creating a rigorous basis for its use in finance.
\vspace{0.2cm} \\ Within quantitative finance, interacting particle systems and Sequential Monte Carlo (SMC) methods have already been applied to derivative pricing \cite{jasra2011sequential, rambharat2010sequential, Shevchenko_2016}. Adaptive Multilevel Splitting (AMS) can be understood as a direct evolution of this idea \cite{c1}: instead of resampling based on fixed survival sets or generic particle weights, AMS works with a scalar importance function on the whole trajectory and introduces an adaptive sequence of levels determined by the empirical distribution of the particles. At each iteration, a fixed fraction of the least-performing particles is killed and replaced by clones of better-performing ones, so that the intermediate levels separating typical from rare trajectories are constructed on the fly. 
\vspace{0.2cm} \\ The contributions of this study are fourfold. First, an AMS adaptation is introduced for binary option pricing under Black–Scholes and Heston dynamics \cite{due, e0f45016-f730-320a-bedc-d34f406805b2}, addressing the specific challenges of risk-neutral valuation and financial time series. Second, the sensitivity of the estimator to parameter choices, including the number of trajectories and resampling rates, is analyzed. Third, numerical experiments compare AMS to standard Monte Carlo, showing substantial gains for deep out-of-the-money options. Fourth, an open source Rcpp implementation is released, supporting Euler, Milstein, and Andersen discretizations \cite{BallyTalay1996, HighamMaoSzpruch2013, Andersen2007}, two importance functions, and six binary option variants, offering a flexible toolkit for rare event simulation in derivatives pricing. A fully documented and publicly available implementation is provided in the R package amsSim \cite{mio}, which has been released on CRAN and includes all the algorithms and numerical experiments presented in this paper.
\vspace{0.2cm} \\ The paper is structured as follows. Section 2 reviews the background on SDE discretization, binary option pricing, and AMS methodology. Section 3 presents the limits of classical variance reduction techniques. Section 4 illustrates the adapted AMS algorithm, establishes its theoretical properties, details the numerical implementation, and reports results against benchmark approaches. Section 5 concludes with a summary of findings and directions for future research.

\section{Research methodology}
\subsection{Stochastic differential equation models}

Numerical experiments are conducted under two standard models for asset price dynamics: the Black–Scholes model \cite{due} and the Heston model \cite{e0f45016-f730-320a-bedc-d34f406805b2}. These frameworks allow assessment of the robustness of the AMS approach across different model complexities.
\vspace{0.2cm} \\ For the Black–Scholes case the exact solution, obtained via logarithmic transformation, removes discretization error \cite{BOYLE1977323, glasserman2004monte}:
\begin{equation}
S_{k+1} = S_k \exp\left[\left(r - \tfrac{\sigma^2}{2}\right)\Delta t + \sigma \Delta W_k\right].
\end{equation}
For the Heston model the variance process requires a scheme that preserves positivity and avoids bias. The quadratic–exponential (QE) method of Andersen \cite{Andersen2007} is employed, the standard approach for accurate Heston simulation. It matches the first two conditional moments of $V_{t+\Delta t}\,|\,V_t$ and selects the update regime according to
\[
\psi \le \psi_c:\quad
V_{t+\Delta t} = a(b + Z)^2,\qquad Z \sim \mathcal{N}(0,1),
\]
\[
\psi > \psi_c:\quad
V_{t+\Delta t} =
\begin{cases}
0 & \text{with probability } p=\dfrac{\psi-1}{\psi+1},\\[4pt]
\beta^{-1}\log\!\left(\dfrac{1-p}{1-U}\right) & \text{with probability } 1-p,
\end{cases}
\]
where $U \sim \text{Uniform}(0,1)$ and $\beta=(1-p)/m$.
\vspace{0.2cm} \\ The asset price is then updated as
\begin{equation}
S_{t+\Delta t} = S_t \exp\!\left[r\Delta t + K_0 + K_1V_t + K_2V_{t+\Delta t} + \sqrt{K_3V_t + K_4V_{t+\Delta t}}\;\epsilon\right],
\end{equation}
with $\epsilon \sim \mathcal{N}(0,1)$. The coefficients $\{K_0,\dots,K_4\}$ and the parameters $a$, $b$, and $\psi$ are given explicitly in \cite{Andersen2007}.
\vspace{0.2cm} \\ This construction preserves the positivity of variance and yields accurate joint dynamics, making it the reference scheme for Heston simulations in rare event pricing.

\subsection{Binary option pricing}
Binary options are derivatives with discontinuous payoffs that activate when the underlying asset satisfies a prescribed condition at maturity or along the path. Three contracts are considered:
\begin{itemize}
    \item \textbf{digital call:} $\;\text{Payoff} = \mathbf{1}_{\{S_T > K\}}$
    \item \textbf{Asian digital call:} $\;\text{Payoff} = \mathbf{1}_{\left\{\tfrac{1}{m}\sum_{t=1}^{m} S_{t} > K\right\}}$
    \item \textbf{up-and-in barrier digital call:} $\;\text{Payoff} = \mathbf{1}_{\left\{\max_{0 \le t \le T} S_t > K_L\right\}}$
\end{itemize}
The discontinuous structure makes pricing highly sensitive to small path variations and leads to large variance under standard Monte Carlo. Difficulties intensify in rare-event regimes, for example, deep out-of-the-money strikes, where the target probability $\mathbb{P}(A)$ is extremely small and crude Monte Carlo requires a prohibitive number of samples. These features make binary options an effective stress test for adaptive multilevel splitting: AMS reallocates computational effort toward trajectories likely to activate the payoff condition, and the payoff itself corresponds to the estimation of a probability, making the method directly and rigorously applicable.

\subsection{Adaptive multilevel splitting (AMS)}
\label{subsec:ams_pseudocode}
Adaptive multilevel splitting (AMS) \cite{Cérou27022007, c1} is a variance reduction method for estimating the probability of rare events. Instead of relying on brute–force Monte Carlo, which wastes almost all trajectories on paths that never approach the rare set, AMS iteratively concentrates simulation effort on trajectories that make progress towards the event of interest.
\vspace{0.2cm}\\
Let $\{X_t\}_{t\ge 0}$ be a Markov process with initial distribution $\eta_0$, and let
\[
p = \mathbb{P}\big(X_\tau \in D\big)
\]
be the probability that the process hits a rare set $D$ at a stopping time $\tau$.  
AMS requires three main ingredients.
\paragraph{Score function, rare level and trajectory score.}
An importance or score function $\xi:\mathbb{R}^d\to\mathbb{R}$ quantifies the progress of the process towards the rare set $D$.  
We fix a rare level $L_{\max}\in\mathbb{R}$ and assume the relaxed condition
\[
x\in D \;\Rightarrow\; \xi(x)\ge L_{\max},
\]
which is sufficient to preserve unbiasedness \cite{brehier2016unbiasedness}.  
Equality between the two sets is not required, although a score function that is more closely aligned with $D$ leads to a lower variance.  
For a complete trajectory $X=(X_t)_{t\in[0,\tau_f]}$ we then define its score as
\begin{equation}
I(X) \;=\; \sup_{t\in[0,\tau_f]} \xi(X_t),
\label{eq:trajectory-score}
\end{equation}
namely the maximal level reached with respect to the importance function.
\paragraph{Population size and killing rate.}
Second, the algorithm works with a population of $N$ replicas and a killing parameter $K$ with $1\le K < N$.  
At each iteration the $K$ worst replicas are removed and replaced by clones of better–performing replicas.  
In practice, choosing $K\le N/2$ preserves enough diversity in the population. 
\medskip \\
The AMS mechanism can be understood by considering a process that must reach a high level $L_{\max}$.  \\
Rather than simulating many independent trajectories from $X_0$ and counting only the few that cross $L_{\max}$, AMS repeatedly:
(i) discards the $K$ trajectories with smallest score, and  
(ii) replaces them with clones of better trajectories that have already crossed an intermediate level.  \\
The clones are restarted from the \emph{first crossing time} of the current level and then resimulated onwards with fresh randomness.  
In this way the whole population is gradually pushed towards the rare region.
\begin{figure}[h]
    \centering
    \includegraphics[width=0.8\textwidth]{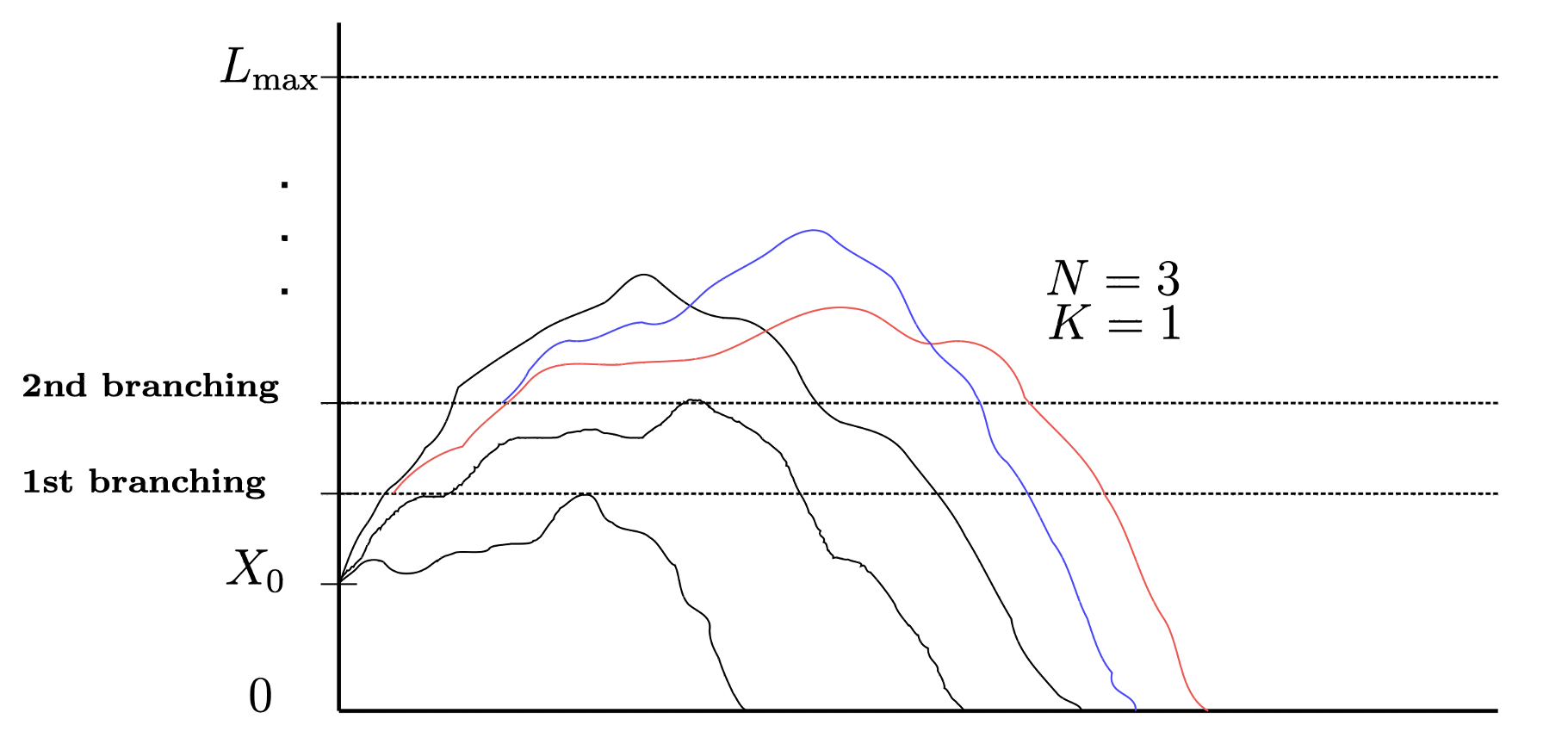}
    \caption{\textit{Illustration of the first two iterations of the AMS algorithm with population size $N=3$ and killing parameter $K=1$.  
    The horizontal dashed lines represent successive adaptive levels $Z^{(1)}$ and $Z^{(2)}$, while the top dashed line denotes the rare level $L_{\max}$.  
    All trajectories start from the common initial point $X_0$.  
    At the first branching level, the trajectory with the lowest score is \emph{killed}; one of the two better trajectories is \emph{cloned} and its clone is restarted from the time at which the original path first crosses $Z^{(1)}$, after which it is resimulated using new random increments (coloured path).  
    The same procedure is repeated at the second branching level.  
    The figure shows how the algorithm progressively reallocates computational effort to trajectories that move closer to the rare-event level $L_{\max}$.}}
    \label{fig:AMS.png}
\end{figure} \newpage
\paragraph{Algorithm description.}
We now formalise the procedure underlying Figure~\ref{fig:AMS.png}.  
Let $X^j=(X_t^j)_{t\in[0,\tau_f]}$, $j=1,\dots,N$, denote $N$ independent replicas of the Markov chain, all started outside $D$.  
Define their scores $S^j = I(X^j)$ using \eqref{eq:trajectory-score}, and let $S^{(1)} \le \dots \le S^{(N)}$ be the ordered scores.  
The algorithm maintains a weight $W$, initialised at $W_0=1$, and proceeds iteratively as follows:

\begin{enumerate}
    \item \textbf{Level selection.}  
    At iteration $q$, compute the order statistics $\{S^{(j)}\}_{j=1}^N$ and set
    \[
        Z_q = S^{(K)},
    \]
    the $K$-th order statistic.  
    The level $Z_q$ corresponds to the horizontal dashed line labelled ``$q$-th branching'' in Figure~\ref{fig:AMS.png}.

    \item \textbf{Stopping criterion.}  
    If $Z_q \ge L_{\max}$, or if $S^{(1)} = \dots = S^{(N)}$, terminate the procedure.

    \item \textbf{Killing and cloning.}  
    Identify the index set of the $K$ worst replicas,
    \[
        \mathcal{K}_q = \{ j : S^j \le Z_q\},
    \]
    and the set of survivors $\mathcal{S}_q = \{1,\dots,N\}\setminus \mathcal{K}_q$.  
    For each $j \in \mathcal{K}_q$:
    \begin{enumerate}
        \item Draw at random an index $i \in \mathcal{S}_q$. 
        \item Let
        \[
            T_i(Z_q) = \inf\{ t \le \tau_f : \xi(X_t^i) \ge Z_q \}
        \]
        be the first crossing time of the current level $Z_q$ by replica $i$.
        \item Construct a new trajectory $\tilde X^j$ by setting
        \[
            \tilde X_t^j = X_t^i \quad \text{for } 0 \le t \le T_i(Z_q),
        \]
        and then simulating fresh dynamics for $t > T_i(Z_q)$ starting from the state $X_{T_i(Z_q)}^i$.  
        In Figure~\ref{fig:AMS.png} this corresponds to the coloured trajectory that coincides with the parent path up to the branching level and then diverges.
    \end{enumerate}

    \item \textbf{Weight update.}  
    Update the common weight as
    \[
        W \;\leftarrow\; \frac{N-K}{N}\,W.
    \]
\end{enumerate}
\noindent Operatively, the adaptive multilevel splitting (AMS) algorithm proceeds as detailed in Algorithm~\ref{alg:AMS}.
\begin{algorithm}[H]
\caption{Adaptive multilevel splitting (AMS)}
\label{alg:AMS}
\begin{algorithmic}[1]
\REQUIRE Population size $N$, killing parameter $K$, importance function $\xi$, rare level $L_{\max}$.
\STATE Generate initial trajectories $\{X^j\}_{j=1}^N$ up to their stopping times $\tau_j$.
\STATE Compute initial scores $S_j \leftarrow I(X^j)$ for each trajectory $j$.
\STATE Sort the scores $\{S_j\}_{j=1,\dots,N}$ as $S_{(1)} \le S_{(2)} \le \dots \le S_{(N)}$.
\STATE Set $Z \leftarrow S_{(K)}$, iteration counter $q \leftarrow 0$ and weight $W \leftarrow 1$.
\WHILE{$Z < L_{\max}$}
  \STATE Let $\mathcal{I} = \{1,\dots,N\}$ be the current index set.
  \STATE Define the set of candidate survivors $\mathcal{S} = \{ j \in \mathcal{I} : S_j > Z \}$.
  \STATE Define the set of candidates to be killed $\mathcal{C} = \{ j \in \mathcal{I} : S_j \le Z \}$.
  \STATE Randomly choose a subset $\mathcal{K}_q \subseteq \mathcal{C}$ of size $K$.
  \STATE Set $\mathcal{S}_q \leftarrow \mathcal{I} \setminus \mathcal{K}_q$ (survivors at iteration $q$).
  \FORALL{$j \in \mathcal{K}_q$}
     \STATE Randomly select a parent index $i \in \mathcal{S}_q$.
     \STATE Compute the first crossing time
         $T_i(Z) \leftarrow \inf\{ t \le \tau_i : \xi(X_t^i) \ge Z \}$.
     \STATE Construct a new trajectory $\tilde X^j$ by setting
       $\tilde X_t^j \leftarrow X_t^i$ for $0 \le t \le T_i(Z)$
       and then resimulating forward for $t > T_i(Z)$ up to its stopping time $\tau_j$,
       using fresh randomness.
     \STATE Replace $X^j \leftarrow \tilde X^j$.
  \ENDFOR
  \STATE Update the scores $S_j \leftarrow I(X^j)$ for all $j=1,\dots,N$.
  \STATE Sort the scores $\{S_j\}$ as $S_{(1)} \le \dots \le S_{(N)}$ and set $Z \leftarrow S_{(K)}$.
  \STATE Update the weight $W \leftarrow \frac{N-K}{N}\,W$.
\ENDWHILE
\STATE Compute the final AMS estimator:
\[
  \hat{p}_{\mathrm{AMS}}
  =
  W \times \frac{1}{N}\sum_{j=1}^{N}\mathbf{1}_{\{X^j \in D\}}.
\]
\end{algorithmic}
\end{algorithm}
\subsubsection{Theoretical properties of AMS}
\paragraph{Well posedness and termination.}  
Let $X=(X_t)_{t\ge 0}$ be a Markov process with importance function $\xi$ and rare set $D$.  
For fixed population size $N$ and killing parameter $K\in\{1,\dots,N-1\}$, AMS is well posed: at each iteration the cutting level $Z$ is an order statistic $Z = S^{(K)}$, and under standard regularity assumptions on $X$ (Feller property), on $\xi$ (continuity), and on the entrance into $D$ (strict entrance condition), the algorithm terminates almost surely after a finite number of iterations \cite{c1}.

\paragraph{Unbiasedness.}  
The AMS estimator
\begin{equation}
\hat p_{\mathrm{AMS}}
=
\Big(\tfrac{N-K}{N}\Big)^{Q}\,
\frac{1}{N}\sum_{j=1}^{N}\mathbf{1}_{\{X^{j}\in D\}}
\end{equation}
is unbiased for any admissible choice of $\xi$ and $K$, where $Q$ denotes the (random) number of iterations.  
It is sufficient that the rare set is contained in a super-level set of the score,
\[
D \subset \{ x : \xi(x) \ge L_{\max} \},
\]
without requiring the stronger equivalence $\xi(x)\ge L_{\max}\iff x\in D$ \cite{brehier2016unbiasedness,10.1093/biomet/asy028,c1}.  
Unbiasedness extends to unnormalised measures
\[
\gamma(\varphi)
\;=\;
\mathbb{E}\big[\varphi(X_\tau)\,\mathbf{1}_{\{X_\tau\in D\}}\big],
\]
for bounded test functions $\varphi$.  \\
A crucial technical point is that randomised cloning and correct handling of times in the scores are essential to preserve unbiasedness. In practice, any deterministic rule for selecting which trajectories to kill or clone breaks exchangeability and introduces bias.
\paragraph{LLN and CLT.}  
Let $\gamma^{(N)}(\varphi)$ denote the AMS approximation of $\gamma(\varphi)$ based on $N$ replicas.  
A law of large numbers holds:
\[
\gamma^{(N)}(\varphi)\ \xrightarrow[N\to\infty]{\mathbb{P}}\ \gamma(\varphi).
\]
Under additional regularity assumptions, a central limit theorem is available,
\[
\sqrt{N}\Big(\gamma^{(N)}(\varphi)-\gamma(\varphi)\Big)\ \Rightarrow\ \mathcal{N}\big(0,\sigma^2(\varphi)\big),
\]
where the asymptotic variance $\sigma^2(\varphi)$ can be characterised via the associated Fleming–Viot particle system \cite{c1,doi:10.1137/18M1187477}.  
In particular, for $K=1$ and a rare event with probability $p$,
\[
\sqrt{N}(\hat p_{\mathrm{AMS}}-p)\ \Rightarrow\ \mathcal{N}(0,\sigma^2),
\qquad
-p^{2}\log p \ \le\ \sigma^{2} \ \le\ 2p(1-p).
\]
A general CLT for $K>1$ remains an open problem; however, both theoretical results in simplified settings and numerical evidence indicate the usual $N^{-1/2}$ scaling, with an asymptotic variance comparable to that of Sequential Monte Carlo methods.
\paragraph{Role of the importance function.}  
While the unbiasedness of $\hat p_{\mathrm{AMS}}$ does not depend on the particular choice of the importance function $\xi$, the variance of the estimator is strongly affected by it. In particular, if $\xi$ is poorly adapted to the geometry of the rare event (for example, in analogy with a badly chosen importance sampling density), the resulting estimator may exhibit large variance and therefore a much slower convergence to the true value. In practice, variance is controlled by experimentally comparing alternative definitions of $\xi$ and by suitably adjusting the population size $N$ and the killing parameter $K$ \cite{c1,brehier2016unbiasedness}.
 
\paragraph{Key advantages.}  
AMS adapts intermediate levels and branching rates on the fly, removing the need for a priori specification as in classical Multilevel Splitting \cite{Kahn1951SplittingParticleTransmission} or Sequential Monte Carlo \cite{fba646e2e09d4f7f93e66f71554f16b7, cerou:inria-00071391}. The algorithm maintains a fixed population size $N$, ensuring robustness, parallel efficiency, and predictable memory use. It provides unbiased estimators for both rare event probabilities and unnormalised measures $\gamma(\varphi)$, enabling straightforward parallelization across independent runs \cite{c1,brehier2016unbiasedness,10.1093/biomet/asy028}.

\section{Theoretical comparison with variance reduction techniques}

\subsection{Antithetic variates: overview and limitations}
Antithetic variates \cite{gentle2009antithetic} reduce variance by pairing negatively correlated samples. In option pricing this corresponds to simulating each path together with its reflection obtained by negating Brownian increments. For monotone payoffs the estimator variance decreases, with a theoretical maximum reduction by a factor of two.  
\vspace{0.2cm} \\
For binary options with probabilities as small as $10^{-6}$, a $2\times$ gain is negligible relative to the computational burden.
\subsection{Control variates: overview and limitations}
Control variates reduce variance by exploiting correlation between the payoff $Y$ and an auxiliary variable $W$ with known expectation. The estimator
\[
\hat\psi_{\mathrm{CV}} = \frac{1}{n}\sum_{i=1}^n \bigl(Y_i - \beta (W_i - \mathbb{E}[W])\bigr)
\]
remains unbiased, with optimal $\beta^*=\mathrm{Cov}(Y,W)/\mathrm{Var}(W)$ yielding
\[
\mathrm{Var}(\hat\psi_{\mathrm{CV}}) = \frac{1}{n}\mathrm{Var}(Y)\bigl(1-\rho^2_{Y,W}\bigr).
\]
Variance reduction is therefore effective only when $W$ is strongly correlated with $Y$.
\vspace{0.2cm} \\
In practice, however, there is no unique, model–independent choice of $W$ that works well across products: an effective control variate must be engineered to closely mimic the structure of the specific payoff under the given dynamics. For example, for arithmetic Asian options one can use the optimized lower-bound proxy of \cite{FusaiKyriakou2016}. This makes the technique very powerful when specialized, but at the cost of poor generalizability.
\vspace{0.2cm} \\
For discontinuous payoffs such as digital or Asian binary options, suitable highly correlated controls are often unavailable in a generic form; without such carefully tailored constructions, the achievable variance reduction remains marginal and the method loses much of its appeal as a general-purpose tool for rare events.

\subsection{Multilevel Monte Carlo: overview and limitations}
Multilevel Monte Carlo (MLMC) \cite{m1,m2} exploits a hierarchy of approximations $X_0,\dots,X_L$ of the same quantity. The telescoping identity
\[
\mathbb{E}[X_L] = \mathbb{E}[X_0] + \sum_{\ell=1}^L \mathbb{E}[X_\ell - X_{\ell-1}]
\]
reduces variance by coupling successive levels with shared randomness. The resulting estimator achieves mean square error $\mathcal{O}(\varepsilon^2)$ at cost $\mathcal{O}(\varepsilon^{-2})$, compared to $\mathcal{O}(\varepsilon^{-3})$ for standard Monte Carlo \cite{m1}.  
\vspace{0.2cm} \\
MLMC is effective for standard option pricing but less suited to rare event estimation. In tail regimes, the variance of inter-level differences decays slowly, limiting efficiency for digital and barrier options. Optimal allocation of samples,
\[
N_\ell \propto \varepsilon^{-2}\sqrt{V_\ell/C_\ell},
\]
depends on variances $V_\ell$ that are themselves costly to estimate and may behave irregularly across levels, especially in rare event settings. These features complicate implementation and reduce the expected efficiency gains.
\subsection{Importance sampling: overview and limitations}

Importance sampling (IS) \cite{tokdar2010importance} estimates $\psi=\mathbb{E}[h(X)]$ by sampling from an alternative density $g$ and reweighting:
\[
\hat{\psi}_g = \frac{1}{n}\sum_{i=1}^n h(Y_i)\frac{f(Y_i)}{g(Y_i)}, \qquad Y_i \sim g.
\]
Efficiency depends on the choice of $g$, with the optimal density proportional to $|h(y)|f(y)$, which is generally unavailable.  
\vspace{0.2cm} \\
A common construction is exponential tilting via Girsanov’s theorem. For Brownian driven models, $g_\theta(y)=e^{\theta y-\psi(\theta)}f(y)$ with cumulant generating function $\psi(\theta)=\log \mathbb{E}[e^{\theta Y}]$. The optimal parameter $\theta^*$ satisfies $\psi'(\theta^*)=a$, where $a$ is the rare event threshold.  
\vspace{0.2cm} \\
In rare event regimes IS becomes unstable. When exercise probabilities are of order $10^{-6}$, the equation $\psi'(\theta)=a$ may lack a solution or yield extreme $\theta^*$, and evaluation of $e^{\theta Y}$ produces flat likelihood landscapes with sporadic spikes. In such cases Newton–Raphson and related solvers fail to converge, and stochastic optimisers are equally unreliable \cite{casella2021choice}. \vspace{0.2cm} \\ 
Two further issues are critical. \\
\textbf{Variance explosion:} an inappropriate choice of $g(y)$ can inflate the estimator’s variance instead of reducing it \cite{tokdar2010importance}. \\
\textbf{Payoff specific design:} effective importance sampling must be tailored to the payoff. Binary calls, binary puts, and Asian options require distinct tilting schemes, and multi-asset payoffs add combinatorial complexity \cite{imps}. 
\vspace{0.2cm}\\
AMS can be interpreted as a non-parametric analogue of IS: it requires only an importance function indicating progress toward the rare set, avoiding explicit tilting densities and unstable root finding, and thus offering broader applicability across option classes.
\section{AMS applications in finance}
Having established the theoretical framework, AMS is now applied to binary option pricing under the Black–Scholes and Heston models. The Markov property of both dynamics makes them directly compatible with AMS, which relies on memoryless trajectories. The method is tested on the three binary contracts of Section~2.2, with efficiency gains most evident for deep out-of-the-money options where standard Monte Carlo becomes infeasible.  
\vspace{0.2cm}\\ For the Black--Scholes model, volatility is fixed at $\sigma = 0.2$.  
For the Heston model, parameters are set to $\rho = -0.5$, $\kappa = 2.0$, $\theta = 0.04$, $\psi = 0.3$, $V_0=0.04$. \\
All performance metrics are averaged over 50 independent runs, obtained by combining results from 5 different initial seeds, each used to generate 10 simulations, ensuring robust statistical confidence in the comparative analysis. \\ 
Tests include: a European digital call under Heston dynamics with strike $2.2$; a path-dependent digital up-and-in barrier call under Heston dynamics with barrier
level $K_L = 2.45$; and Asian digital calls under both Black--Scholes (strike $1.7$) and Heston (strike $1.6$). 
These parameter choices place all contracts in comparable rare-event regimes.
\paragraph{Implementation and hardware.}
All numerical experiments were run on a MacBook Pro equipped with an Apple M2 Pro
processor and 16\,GB of RAM, running macOS. The C++ core of the \texttt{amsSim}
library was compiled via \texttt{Rcpp} using the Apple \texttt{clang} compiler with
\texttt{-O3} optimisation flags and no GPU acceleration. All simulations are strictly
single-threaded: OpenMP and other forms of multi-threading are disabled, so the
reported timings correspond to wall-clock times on a single CPU core.
\subsection{Importance function design}
\label{imp}
AMS performance depends critically on the importance function $\xi$, which steers trajectories toward the rare event set. Two constructions are considered:  

\begin{itemize}
   \item \textbf{Path-based functions.} For European binaries, $\xi_t = S_t$; for Asian binaries, $\xi_t$ is the running average; for barrier binaries, $\xi_t$ is the running maximum or minimum, depending on the barrier type. For puts, the sign is inverted. In all cases, setting $L_{\max}$ equal to the strike (or to $K_L$ for barriers) guarantees $D \subseteq \{\xi > L_{\max}\}$.
   \item \textbf{Analytical approximations.} Black–Scholes digital formulas are used as importance functions,  
   \begin{align}
   \text{Call}_{BS} &= e^{-rT}\Phi(d_2), \\
   \text{Put}_{BS} &= e^{-rT}\Phi(-d_2),
   \end{align}
   with $d_2=\tfrac{\ln(S/K)+(r-\sigma^2/2)T}{\sigma\sqrt{T}}$. At each $t$, $S_t$ (or the running average for Asians, and the running maximum or minimum
for barrier options). is inserted as the spot input, regardless of the underlying model. Although exact only for European binaries under Black–Scholes, this construction captures the curvature of the pricing function and improves guidance toward the rare event region. Here $L_{\max}=0.5$ ensures $D \subseteq \{\xi > L_{\max}\}$.
\end{itemize}
\begin{remark}[Unbiasedness of AMS for digital options]
Let $(S_t)$ follow either the Black-Scholes dynamics
\[
dS_t = r S_t\,dt + \sigma S_t\,dW_t,
\]
or the Heston system
\[
\begin{cases}  
dS_t = r S_t\,dt + \sqrt{V_t}\,S_t\,dW_t^{(1)}, \\[4pt]  
dV_t = \kappa(\theta - V_t)\,dt + \psi \sqrt{V_t}\,dW_t^{(2)},  
\end{cases}
\]
with $(W^{(1)},W^{(2)})$ a correlated Brownian pair. In both cases the state process is Markovian. \\
Let $D$ be the rare–event set corresponding to the digital payoff 
(European, Asian or barrier call or put). For the importance functions $\xi$ introduced in 
Section~\ref{imp}, the sufficient condition
\[
x \in D \;\;\Rightarrow\;\; \xi(x) \ge L_{\max}
\]
of \cite[Theorem 3.2]{brehier2016unbiasedness} holds. Then the AMS estimator of the risk–neutral probability $p=\mathbb{Q}(D)$ is
\[
  \hat{p}_{\mathrm{AMS}}
  =
  \left(\prod_{i=0}^{q}\frac{N-K}{N}\right)
  \times
  \frac{1}{N}\sum_{j=1}^{N}\mathbf{1}_{\{X^{(j)} \in D\}},
\]
where $q$ is the number of iterations required to reach the threshold $L_{\max}$. 
This estimator is unbiased, and the digital option value
\[
\hat V = e^{-rT}\hat p_{\mathrm{AMS}}
\]
is therefore an unbiased estimator of the true price, with the same asymptotic variance properties as in the general AMS framework.
\end{remark}
\subsubsection{Heuristics and diagnostics for importance function design}
From a practical standpoint, the importance function $\xi$ should be chosen so that it is monotonically aligned with the event of interest. Writing the rare event as $D = \{x : G(x) > 0\}$ for a suitable functional $G$ of the path, a natural heuristic is to take $\xi(x)$ as a smooth surrogate of $G(x)$: trajectories closer to $D$ (in the sense of $G(x)$ being larger) should also have larger $\xi(x)$. In our constructions this is enforced explicitly: for digital payoffs we use functionals that increase as the underlying approaches the exercise region (terminal value, running average, running extremum).
\vspace{0.2cm}\\
From a theoretical perspective, a first requirement is that $\xi$ preserves the support of the event, in the sense that
$
x \in D \;\;\Longrightarrow\;\; \xi(x) \ge L_{\max},
$
so that the sufficient condition of \cite[Theorem~3.2]{brehier2016unbiasedness} is satisfied. If this condition were violated, AMS could in principle miss part of $D$ and introduce bias. Even when unbiasedness holds, a poorly informative choice of $\xi$ (for instance, a score that is nearly uncorrelated with $\mathbf{1}_D$ or that does not increase as trajectories move towards $D$) inflates the variance and can lead to worse efficiency than crude Monte Carlo. 

\subsection{Impact of the selection parameter $K$ on algorithm performance}
\label{kkk}
\begin{figure}[h]
    \centering
    \includegraphics[width=0.8\textwidth]{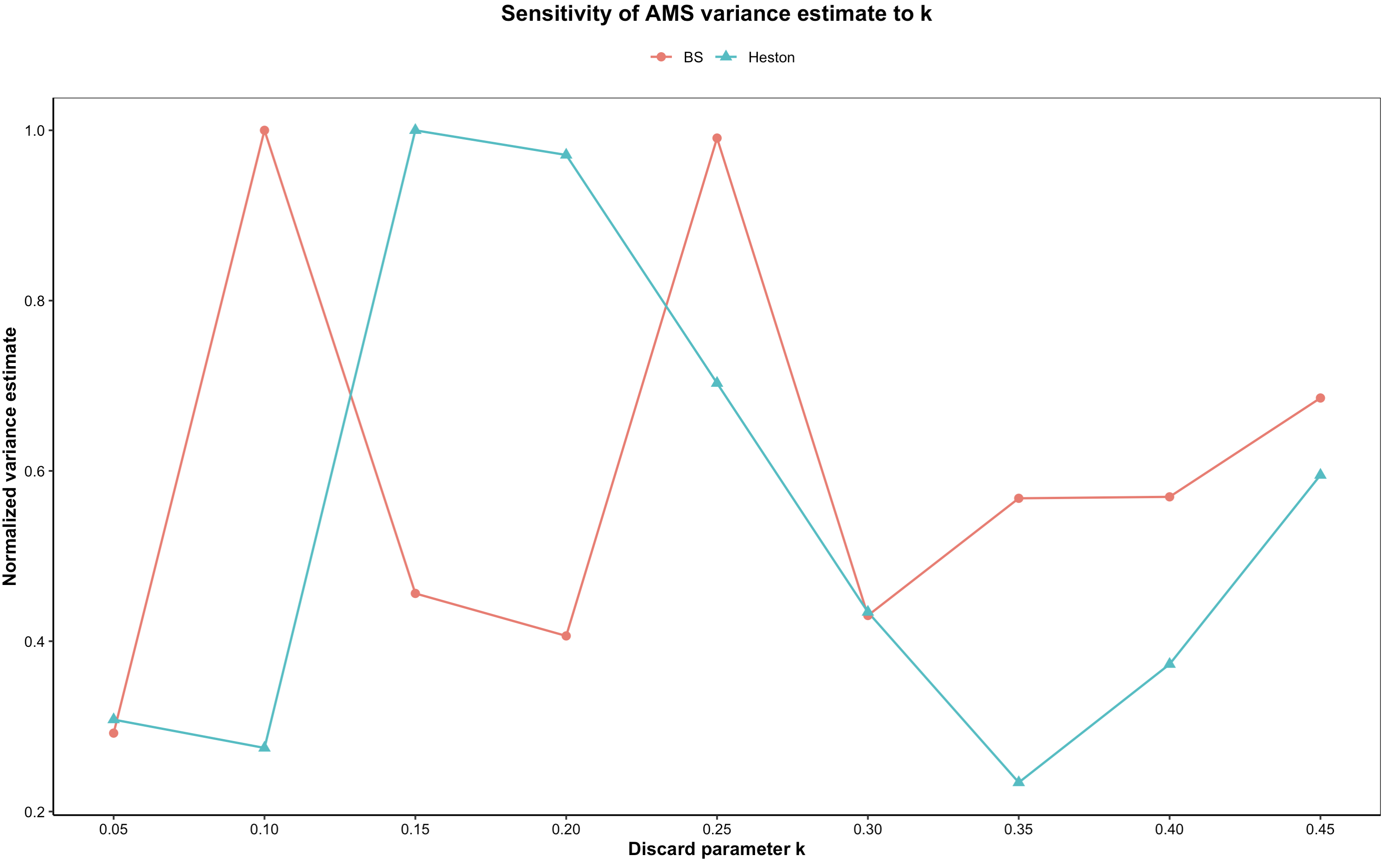}
    \caption{\textit{Normalized variance (vertical axis) as a function of the discard fraction k (horizontal axis).}}
    \label{fig:kkk.png}
\end{figure}

The selection parameter can be expressed either as the absolute number of discarded trajectories $K$, or equivalently as the discard fraction $k = K/N$, where $N$ is the population size. At each iteration exactly $K = kN$ particles (i.e., a fraction $k$ of the population) are killed and replaced by clones. Theoretical results show that, for an optimal importance function, the asymptotic variance is minimised in the limit $K=1$ \cite{c1,doi:10.1137/18M1187477}, but this regime is computationally impractical in realistic settings.
\vspace{0.2cm} \\
In our experiments we fix $N = 50{,}000$ particles and vary the discard fraction $k$ from 5\% to 45\% in steps of 5\%, so that $K = kN$ trajectories are removed at each iteration. We investigate this dependence for a digital call under Heston dynamics and for an Asian digital call under Black--Scholes, both using the path-based importance function described in Section~\ref{imp}.
\vspace{0.2cm} \\
The results are reported in Table \ref{tab:K_time_models} and
Figure \ref{fig:kkk.png}.
\begin{table}[h]
\centering
\caption{Execution time of the AMS algorithm for different rejection rates $k$ under two option pricing settings.}
\begin{tabular}{l r r}
\hline
$k$ & Time (Digital, Heston)  & Time (Digital Asian, Black-Scholes) \\
\hline
0.05 & 35.86  & 25.23 \\
0.10 & 20.02 & 13.46 \\
0.15 &  13.97 &  9.36 \\
0.20 & 11.41 &  7.36 \\
0.25 & 10.20 &  6.08 \\
0.30 & 8.96 & 5.19\\
0.35 & 7.8 & 4.53 \\
0.40 & 7.38  &  4.03 \\
0.45 & 6.7 &  3.63\\
\hline
\end{tabular}
\label{tab:K_time_models}
\end{table}
 \newpage
\noindent In the figure, gray markers correspond to the standard digital call option, while red markers represent the Asian digital call option. \\ Results confirm the trade-off: small $K$ requires more iterations and substantially longer runtime (up to 30 seconds in the Heston case). Estimator quality, however, shows no clear monotonic dependence on K; for these options, performance remains stable across the tested range. 
\subsection{Impact of the number of trajectories N on algorithm performance}
The particle count $N$ directly affects AMS performance. Larger $N$ reduces estimator variance but increases runtime due to higher simulation and sorting costs. Theoretical analysis shows complexity of order $N(-\log(p))\log(N)$, accounting for the sorting step and the generation of one new trajectory per iteration \cite{c1,doi:10.1137/18M1187477}.  
\vspace{0.2cm} \\
Numerical experiments under both Black–Scholes and Heston models, using the options of Section~\ref{kkk}, confirm this trade-off. All tests use $K=0.45$ and the path-based importance function. 
\begin{table}[h]
\centering
\caption{Execution time of the AMS algorithm as a function of the number of trajectories $N$ under two option pricing settings.}
\begin{tabular}{l r r}
\hline
$N$ & Time (Digital, Heston)  & Time (Digital Asian, Black-Scholes)  \\
\hline
 50000   &  5.88 &  3.96 \\
 70000   &  8.79 &  6.08 \\
90000   & 11.48 &  7.94\\
110000   & 14.03 &  9.66 \\
130000   & 16.84 & 11.62 \\
150000   & 19.44 & 13.37 \\
170000   & 22.2 & 15.12 \\
190000   & 25.06 & 17.19 \\
210000   & 28.03 & 19.05 \\
\hline
\end{tabular}
\label{tab:N_time_models}
\end{table} \vspace{0.2cm} \\  
\noindent As reported in Table~\ref{tab:N_time_models}, the computational cost increases steadily with $N$, as expected. 
To compare with the predicted $N(-\log p)\log N$ complexity, we consider the ratio between the observed execution time and $N\log N(-\log p)$. \\
In our experimental settings for the Heston digital option this ratio is approximately $1.1\times 10^{-5}(-\log p)$ across all tested values of $N$, while for the Black--Scholes Asian digital option it is about $7.5\times 10^{-6}(-\log p)$, indicating an essentially constant prefactor and thus supporting the theoretical complexity analysis.
\vspace{0.2cm} \\
These results highlight the inherent balance between variance reduction and runtime when tuning $N$ for AMS in option pricing applications.
\subsection{Analysis of option pricing results}
\label{res}
\begin{figure}[H]
    \centering
    \makebox[\textwidth][c]{%
        \includegraphics[width=1\textwidth]{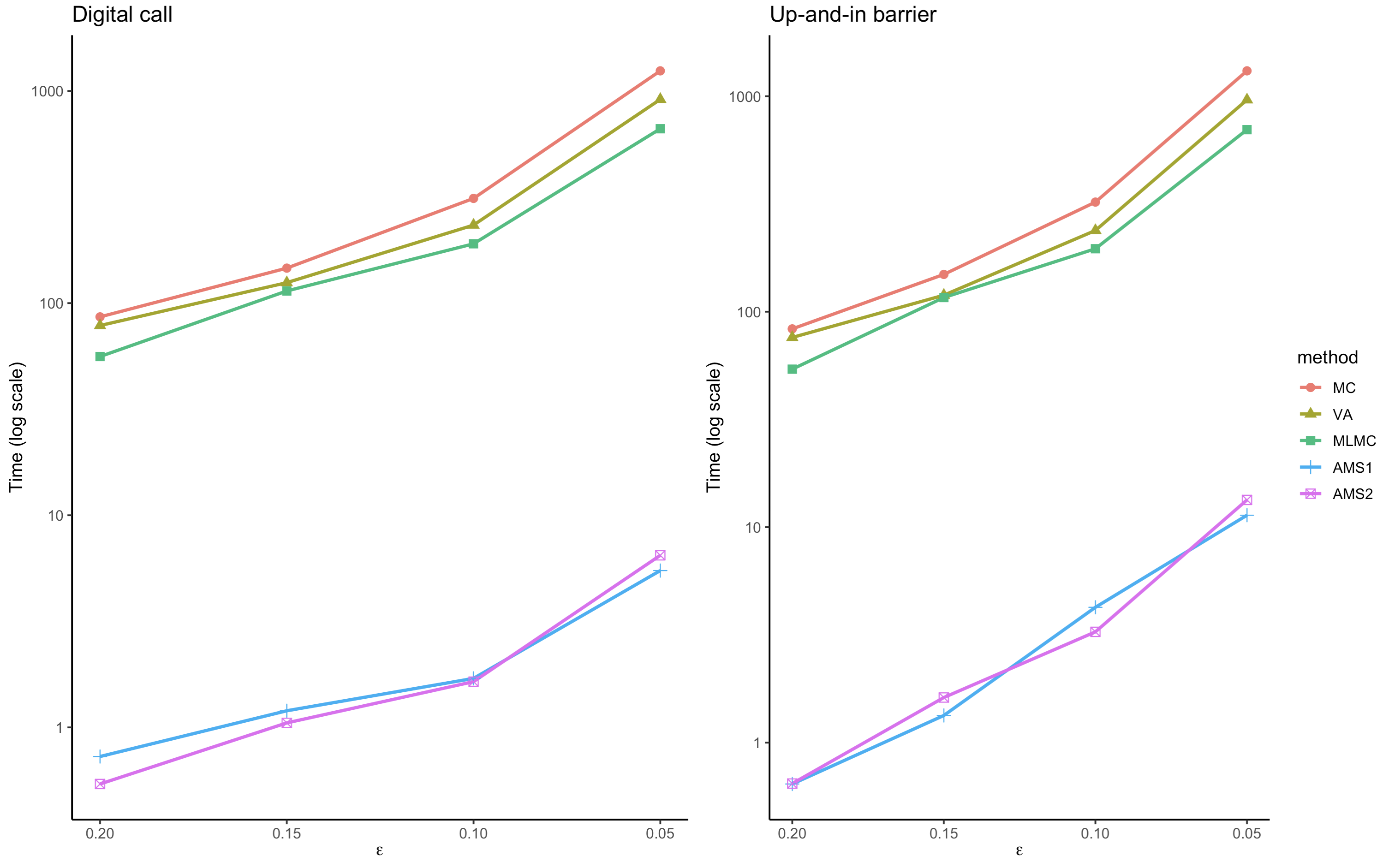}%
    }
    \caption{\textit{Computational time (log scale) as a function of relative accuracy for different simulation methods for the Heston digital call and digital up-and-in barrier call; numerical values are reported in Tables~\ref{tab:heston_digital_call}, \ref{tab:heston_digital_put}}}
    \label{fig:g1}
\end{figure}

\begin{figure}[H]
    \centering
    \makebox[\textwidth][c]{%
        \includegraphics[width=1\textwidth]{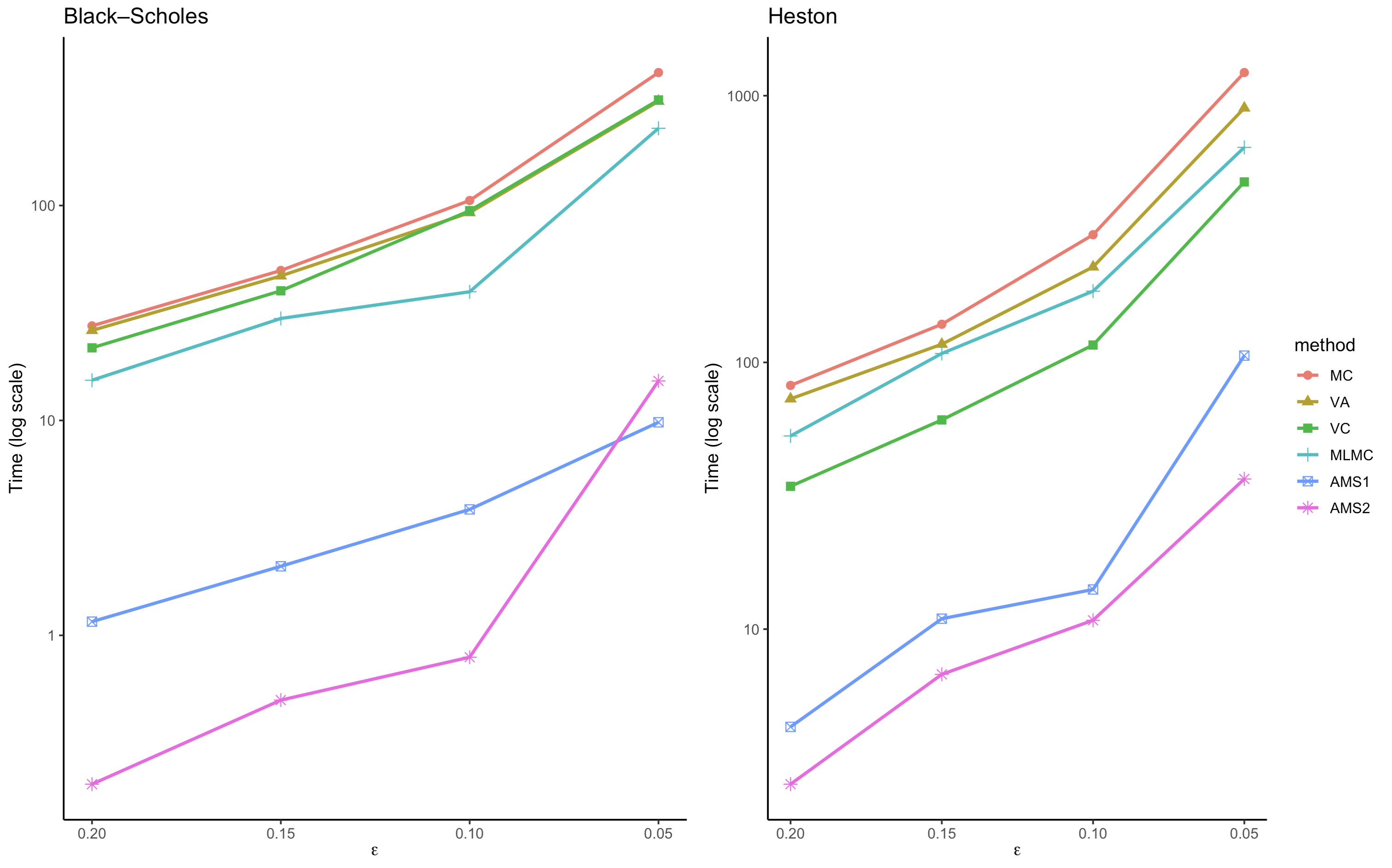}%
    }
    \caption{\textit{Computational time (log scale) as a function of relative accuracy for different simulation methods for the Black-Scholes and Heston digital Asian call; numerical values are reported in Tables~\ref{tab:bs_Asian_digital_call}, \ref{tab:heston_Asian_digital_call}}}
    \label{fig:g2}
\end{figure}

Standard Monte Carlo serves as the primary benchmark for all contracts, together with
antithetic variates, multilevel Monte Carlo (MLMC), and adaptive multilevel splitting
(AMS) using two different importance functions. These constitute the core comparison
set across all experiments. In addition, for specific payoffs further specialised methods
are included when appropriate, enabling a contract-by-contract performance assessment.
\vspace{0.2cm} \\
Test cases focus on deep out-of-the-money contracts, with exercise probabilities down
to order $10^{-6}$, where AMS is expected to deliver its largest efficiency gains.
In our experiments, substantial speed-ups over crude Monte Carlo only start to appear once the probability drops below roughly $10^{-4}$; for less extreme strikes, standard Monte Carlo remains competitive and AMS offers at best modest improvements. These moderately rare events are therefore not the primary target of the present study.
\vspace{0.2cm} \\
Performance is evaluated in terms of \textbf{computational time} (horizontal axis) and \textbf{relative accuracy} (vertical axis), defined as $
  \frac{\sqrt{\mathrm{Var}}}{\mathrm{Mean}}$, 
with the mean and variance estimated over $50$ independent runs. Figures~\ref{fig:g1} and \ref{fig:g2} summarise these results: each curve represents one simulation method for a given contract, and each point on a curve corresponds to a different computational budget. Times are shown on a logarithmic scale, so vertical separations between curves can be interpreted directly as orders of magnitude in speed-up. The underlying numerical values used to generate these plots are reported in Tables~\ref{tab:heston_digital_call}–\ref{tab:heston_Asian_digital_call}.
\vspace{0.2cm} \\
The discard fraction is set to $k=0.45$ (so that $K=0.45N$ trajectories are resampled at each iteration), as smaller values did not yield systematic variance reduction for the same total computational budget (Section~\ref{kkk}).
\vspace{0.2cm} \\
We denote by MCA Monte Carlo with antithetic variates, and by MCV Monte Carlo with the optimised lower-bound control variate of Fusai and Kyriakou \cite{FusaiKyriakou2016}.

\FloatBarrier
\noindent
Figures~\ref{fig:g1} and \ref{fig:g2} show that, across all contracts considered, the AMS curves lie strictly below the corresponding Monte Carlo and MLMC curves over the range of relative accuracies of interest. In other words, for any target relative error in the plots, AMS attains that accuracy at a lower computational cost. This visual separation is most pronounced in the rare-event regime, where standard Monte Carlo must simulate a very large number of paths before observing enough exercise events.\\
In Figure~\ref{fig:g2}, an additional benchmark is included: the control variate method
derived from the optimised lower bound introduced by \cite{FusaiKyriakou2016}. This
provides a further point of comparison, confirming that even when specialised variance–reduction
techniques tailored to the payoff structure are incorporated, \textsc{AMS} remains
the most computationally efficient method across the accuracy levels examined.
\vspace{0.2cm} \\
\textbf{Computational time reduction.}  
For the standard European digital option, \textsc{AMS} achieves speedups exceeding
$100\times$ relative to standard Monte Carlo, peaking above $200\times$ around the
$5\%$ relative accuracy level. The gap between the \textsc{AMS} and MLMC curves remains
close to two orders of magnitude throughout this regime. A similarly pronounced effect
is observed for the digital up-and-in barrier option under Heston: \textsc{AMS} retains
speedups consistently above $100\times$ even when compared with the best-performing
baseline, namely MLMC.
\vspace{0.2cm}\\
For Asian digital options under Black--Scholes, the performance gains are still
substantial. Relative to standard Monte Carlo, \textsc{AMS} improves efficiency by
approximately $25$–$40\times$, while gains over MLMC remain in the $15$–$20\times$
range. Under Heston dynamics the advantage persists, particularly for the second
importance function: \textsc{AMS2} achieves improvements of $30$–$35\times$ over
classical Monte Carlo, and outperforms the specialised control–variate benchmark by
more than one order of magnitude.
\vspace{0.2cm} \\
\textbf{Role of the importance function.}  
The two importance functions produce curves that are close in Figures~\ref{fig:g1} and \ref{fig:g2}, indicating that AMS is not overly sensitive to this modelling choice in the configurations tested. Nevertheless, the Black-Scholes based importance function (AMS2) tends to yield slightly lower times at a fixed relative accuracy in several panels, consistent with the intuition that the $\Phi(d_2)$ term provides a smoother and more informative guidance towards the rare-event region.
\vspace{0.2cm} \\
\textbf{Consistency across option types.}  
The patterns observed in Figure~\ref{fig:g1} for the digital and up-and-in barrier options are
essentially reproduced in Figure~\ref{fig:g2} for the Asian digital case: the relative
ordering of the methods remains unchanged, and \textsc{AMS} consistently emerges as the
most efficient approach in all deep out-of-the-money configurations, for both European
and Asian binaries and under both Black--Scholes and Heston dynamics. This indicates
that the efficiency gains achieved by \textsc{AMS} are robust with respect to the payoff
specification as well as to the underlying volatility model.
\vspace{0.2cm}\\
Taken together, Figures~\ref{fig:g1} and \ref{fig:g2} demonstrate that \textsc{AMS}
produces unbiased estimates while delivering substantial computational savings relative
to standard Monte Carlo, and remains competitive against advanced variance–reduction
techniques such as MLMC or control variates. The advantage is particularly pronounced
in the rare-event regime that motivates the use of \textsc{AMS}, where conventional
methods become prohibitively expensive.
\subsubsection{Theoretical analysis of Greek estimation in adaptive frameworks}
Estimating sensitivities (Greeks) within the AMS framework presents specific challenges related to the adaptive nature of the algorithm. This section contrasts the Finite Difference approach with direct differentiation methods, highlighting the structural bias affecting the latter.
\vspace{0.2cm}\\
\textbf{Finite Difference benchmark.} The Finite Difference (FD) method serves as a robust baseline. For a parameter $\theta$ and perturbation $h$, the estimator is \[\partial_\theta V \approx \frac{\hat{V}(\theta + h) - \hat{V}(\theta)}{h} \] 
Although FD typically suffers from variance inflation as $h \to 0$, the variance reduction achieved by AMS on the price estimator $\hat{V}$ mitigates this instability. This makes FD a viable strategy without requiring prohibitive computational budgets.
\vspace{0.2cm}\\
\textbf{Bias in pathwise differentiation.} The conditional pathwise method \cite{glasserman2004monte} was implemented to address the non-differentiability of binary payoffs. This technique smoothens the discontinuity by conditioning on the state of the process just prior to maturity (at $T-\Delta t$) and integrating out the final Brownian increment. This effectively replaces the indicator $\mathbf{1}_{\{S_T > K\}}$ with the conditional probability of exercise, a smooth function that permits the interchange of differentiation and expectation.
\\
While this approach yields substantial variance reduction, it exhibits a systematic negative bias in the AMS setting. The bias stems from the dependence of the AMS weight on model parameters. Considering the estimator $\hat{V}_{\text{AMS}} = W_{\text{AMS}} \times \hat{Y}_N$, the total derivative requires the product rule:
\begin{equation}
    \frac{\partial \hat{V}_{\text{AMS}}}{\partial \theta} = W_{\text{AMS}} \frac{\partial \hat{Y}_N}{\partial \theta} + \hat{Y}_N \frac{\partial W_{\text{AMS}}}{\partial \theta}.
\end{equation}
Standard pathwise schemes operate under a ``frozen weight'' approximation, computing only the first term. However, the stopping time $Q$ in AMS is parameter-dependent: for a Call option, an increase in $S_0$ reduces the expected iterations ($\partial Q / \partial S_0 < 0$), implying $\partial W_{\text{AMS}} / \partial S_0 > 0$. Neglecting this term causes the observed underestimation.
\vspace{0.2cm}\\
\textbf{Computational trade-off.} This issue mirrors the path degeneracy problem documented in Sequential Monte Carlo (SMC) literature \cite{jasra2011sequential}. Jasra and Del Moral demonstrated that the resampling mechanism, central to both SMC and AMS, introduces discrete updates that disrupt the smooth dependence of trajectories on parameters. Consequently, standard gradient estimators fail to account for the changes in particle genealogy induced by parameter perturbations. 
\\
Retrieving unbiased gradients via the Likelihood Ratio Method (LRM) necessitates integrating over all possible ancestral lineages to capture the weight sensitivity, imposing a computational complexity of $\mathcal{O}(N^2)$. For the large population sizes required in rare event simulation ($N \sim 10^5$), such quadratic cost is prohibitive. Conversely, linear complexity approximations exist but suffer from asymptotic bias.
\vspace{0.2cm}\\
Finite Differences represent the most effective trade-off: they avoid the systematic bias of pathwise smoothing and the prohibitive cost of LRM. While subject to discretization error, they implicitly capture the sensitivity of the stopping time $Q$ while maintaining linear $\mathcal{O}(N)$ complexity.

\subsubsection{Numerical stress test}
An extreme scenario is considered to further test AMS. A digital option under Black–Scholes with $S_0=1$, $K=3.5$, $T=1$, and $r=0.03$ has analytical value $2.509\times 10^{-10}$. Only the path-based importance function is used, to avoid embedding model information into $\xi$.  
\vspace{0.2cm} \\
Table~\ref{tab:extreme} reports results for a $10\%$ relative accuracy target. For Monte Carlo, execution time is extrapolated analytically. With  
\[
\epsilon = \frac{\sqrt{\mathrm{Var}(\hat{p})}}{p} = \frac{\sqrt{p(1-p)/N}}{p},
\]
the required $N$ is $(1-p)/(\epsilon^2 p) \approx 4\times 10^{11}$, corresponding to $T_{\mathrm{MC}}\approx 3.2\times 10^6$ seconds ($\sim$888 hours) given $10^6$ paths in 8 seconds.  

\begin{table}[h!]
\centering
\caption{Comparison between Monte Carlo and AMS in the extreme scenario.}
\label{tab:extreme}
\begin{tabular}{c|c|c}
\toprule
& Monte Carlo & AMS  \\
\midrule
Time (s) & 3{,}200{,}000 & 29.979 \\
\bottomrule
\end{tabular}
\end{table}
\noindent AMS attains the target within 30 seconds, confirming its robustness in extreme rare event regimes where standard Monte Carlo is computationally infeasible.
\subsubsection{Multi-asset digital option analysis}
\label{multioption}
To illustrate the flexibility of AMS in a multidimensional setting, we consider
a synthetic three-asset digital option that does not correspond to a traded product and is
not included in the \texttt{amsSim} library. The contract is introduced solely as a
stress test in a multi-asset rare-event scenario.
\vspace{0.2cm}\\ We work under a Black--Scholes framework with three underlying assets
$S^{(1)}, S^{(2)}, S^{(3)}$ following correlated geometric Brownian motions under the
risk–neutral measure $\mathbb{Q}$. The driving Brownian motions have constant pairwise
correlation $\rho = 0.2$. In the numerical experiment we take a common initial price
$S_0^{(i)} = 1$ for all three assets, a constant volatility $\sigma = 0.2$, and a
constant risk-free rate $r = 0.03$. \\ At maturity $T$ we define the arithmetic average
\[
   \bar{S}_T = \frac{1}{3} \bigl( S_T^{(1)} + S_T^{(2)} + S_T^{(3)} \bigr)
\]
and the maximum pairwise terminal price difference
\[
   \Delta_{\max}
   =
   \max \Bigl\{
      \bigl| S_T^{(1)} - S_T^{(2)} \bigr|,
      \bigl| S_T^{(2)} - S_T^{(3)} \bigr|,
      \bigl| S_T^{(1)} - S_T^{(3)} \bigr|
   \Bigr\}.
\]
Given a dispersion threshold $L > 0$ and an average level $K_{\text{avg}} > 0$, the payoff is
\begin{equation}
   \Pi =
   \begin{cases}
      1, & \text{if } \Delta_{\max} > L \text{ and } \bar{S}_T > K_{\text{avg}},\\[4pt]
      0, & \text{otherwise,}
   \end{cases}
   \label{eq:multi_payoff}
\end{equation}
In the parameter setting used here, the maximum price difference must exceed
$L = 1$, while the arithmetic average must satisfy $\bar{S}_T > K_{\text{avg}} = 1.4$,
with $S_0^{(i)} = 1$ for all assets. \\ 
The payoff is path-independent but strongly non-linear and multidimensional, and no closed-form solution is available in the classical Black--Scholes setting. \\ The payoff is deliberately designed to
combine two competing features: on the one hand, it rewards configurations in which the
assets exhibit strong cross-sectional dispersion (two prices moving far above the strike
while the remaining one moves far below it), so that the rare event is driven by extreme
relative movements; on the other hand, an additional condition on the cross-sectional
average imposes a common level constraint that prevents trivial scenarios where the whole
basket drifts to a very low mean.
\vspace{0.2cm}\\ For the present three-asset option we define the score function as
\[
   \xi =
   \Delta_{\max} + \bar{S}_T,
\]
so that trajectories closer to the rare-event region
$\{\Delta_{\max} > L,\; \bar{S}_T > K_{\text{avg}}\}$ receive larger scores. Choosing
$L_{\max}=L+K_{\text{avg}}$ such that
\[
   (\Delta_{\max},\bar{S}_T) \in \{ \Delta_{\max} > L,\; \bar{S}_T > K_{\text{avg}} \}
   \;\Rightarrow\;
   \xi \ge L_{\max}
\]
ensures that the sufficient condition of \cite[Theorem~3.2]{brehier2016unbiasedness}
holds and the AMS estimator remains unbiased. 
\vspace{0.2cm}\\
In the multi-asset setting, the set of viable benchmarks becomes considerably more
restricted. MLMC, although effective for one-dimensional path functionals, does not
admit a straightforward extension to this payoff, whose rare-event structure combines
cross-sectional dispersion and joint level constraints; constructing a suitable hierarchy
and coupling mechanism is non-trivial.\\
Importance sampling (IS) is even more delicate. In high dimensions, an effective change
of measure must simultaneously bias the process towards large relative deviations and a
higher average level, leading to a highly anisotropic rare set. Identifying such a tilt is
problem-specific and unstable. Even cross-entropy methods, often used to automate the
search for an optimal IS density, become difficult to apply: selecting a parametric family
capable of capturing the joint tail behaviour is non-obvious.\\
Due to these challenges, we restrict the comparison to standard Monte Carlo and Monte
Carlo with antithetic variates.\\ The corresponding performance metrics, together with those of the AMS estimator, are reported in Table~\ref{tab:variance_multi}. 
\begin{table}[H]
\centering
\caption{Computational times (in seconds) for different relative accuracy levels in the multi-dimension call experiment.}
\label{tab:variance_multi}
\begin{tabular}{lrrr}
\hline
\textbf{Relative accuracy} & \textbf{MC} & \textbf{MCA} & \textbf{AMS} \\
\hline
0.05 & 23.62 & 19.22 & 0.44 \\
0.03 & 67.89 & 58.93 & 1.25 \\
0.01 & 555.49 & 438.13 & 12.19 \\
\hline
\end{tabular}
\end{table}
\noindent As in the one-dimensional setting, AMS achieves a substantial computational cost reduction for 
comparable variance, confirming its effectiveness also in this 
multidimensional rare-event scenario.
\newpage
\section{Conclusions and future work}
\subsection{Conclusions}
This study establishes adaptive multilevel splitting (AMS) as a computationally superior method for pricing binary options in rare event regimes. Across both Black–Scholes and Heston models, AMS achieves speedups of up to 200 over standard Monte Carlo while maintaining unbiasedness, and consistently outperforms variance reduction baselines such as MLMC and antithetic variates.  
\vspace{0.2cm} \\
To the best of our knowledge, this is the first application of AMS to financial rare event pricing. Benchmarking against the closest variance reduction methods in finance confirms its superior efficiency in deep out-of-the-money regimes, where conventional techniques become computationally infeasible.  
\vspace{0.2cm} \\
The practical implications are significant: AMS renders previously intractable problems feasible, enabling tighter spreads and deeper liquidity for rare event derivatives, with direct relevance for parametric insurance and catastrophe-linked products.  
\vspace{0.2cm} \\
The method also shows strong scalability. Importance functions are simple to construct and adaptable across payoff types, and performance is less sensitive to their specification than in importance sampling. This robustness facilitates deployment in both academic and industry settings.  
\subsection{Future developments}
The success of AMS in binary option pricing suggests several extensions beyond derivatives valuation.  
\vspace{0.2cm} \\
A first direction is risk management, where AMS could improve the computation of tail risk measures. Value-at-Risk (VaR), defined as the loss threshold exceeded with small probability, is a rare event problem. Existing Monte Carlo and importance sampling approaches are widely used \cite{hong2014monte,sun2010asymptotic}; AMS offers the potential for more accurate and efficient estimates, relevant for stress testing and regulatory capital.  
\vspace{0.2cm} \\
A second extension concerns model coverage. Incorporating exotic payoffs and multi-asset structures would broaden applicability, enabling AMS to address higher-dimensional rare event problems and increasing the versatility of the package for quantitative finance.  
\vspace{0.2cm} \\
A third avenue is methodological. Rough volatility models such as Bergomi \cite{gatheral2017rough} pose challenges because fractional Brownian motion violates the Markov property central to AMS. One possible solution is a lifted Markovian approximation embedding the non-Markovian dynamics in a higher-dimensional state space \cite{zhu2021markovian}, potentially extending AMS to this class of models.
\newpage
\section*{Appendix}
\appendix
\section{C++ implementation with R interface via Rcpp}
Prior to this work, no R package provided AMS functionality for financial applications.
To fill this gap, a dedicated implementation was developed in C++ \cite{press2007numerical} with an R interface via Rcpp \cite{eddelbuettel2013seamless}.  
\vspace{0.2cm} \\
The algorithmic structure of AMS, nested loops over splitting levels, trajectory simulation, and resampling, requires extensive floating-point operations and predictable memory access, making compiled code essential. The C++ engine employs pre-allocated trajectory containers, object pooling, vectorized SDE discretization, efficient random number generation, and in-place sorting to minimize memory and copying overhead.  
\vspace{0.2cm} \\
The Rcpp interface exposes all algorithmic parameters and diagnostics within the R environment, while computationally intensive tasks remain in C++. This design combines the usability of R with near native performance, enabling practical deployment of AMS in quantitative finance. The complete implementation is distributed as the R package amsSim \cite{mio}, available on CRAN.

\subsection{Core implementation}

The Rcpp implementation is organized into a set of core functions that handle stochastic simulation, payoff evaluation, importance function construction, and execution of the AMS algorithm (Table~\ref{tab:functions}).  

\begin{table}[htbp]
\centering
\caption{Summary of core functions}
\label{tab:functions}
\begin{tabular}{p{0.3\textwidth}p{0.6\textwidth}}
\toprule
\textbf{Function} & \textbf{Description} \\
\midrule
simulateAMS & Generates Monte Carlo paths. Implements exact Black–Scholes discretization and three Heston schemes: Euler–Maruyama, Milstein, and Andersen’s Quadratic–Exponential.  \\
\addlinespace[0.2cm]
payoff & Evaluates six exotic payoffs: digital call, digital put, Asian digital call, Asian digital put, up-and-in barrier call, and up-and-in barrier put. \\
\addlinespace[0.2cm]
functionAMSCpp & Computes the two importance functions described in Section~\ref{imp}. \\
\addlinespace[0.2cm]
AMS & Executes the full AMS algorithm, integrating path generation, resampling, and weighting. Supports six payoff types and two importance functions. Parameters include strike, $L_{\max}$, and selection fraction $K$. \\
\bottomrule
\end{tabular}
\end{table}
\noindent \textbf{Code availability.} The full implementation, including C++ source files and the R interface, is publicly available at \url{https://github.com/RiccardoGozzo/amsSim}.
\section{Tables underlying the figures}
\begin{table}[h!]
\centering
\caption{Computational times (in seconds) for different relative accuracy levels in the Heston digital call experiment.}
\label{tab:heston_digital_call}
\begin{tabular}{lrrrrr}
\hline
\textbf{Relative accuracy} & \textbf{MC} & \textbf{MCA} & \textbf{MLMC} & \textbf{AMS 1} & \textbf{AMS 2} \\
\hline
0.20 & 86.13 & 78.39 & 56 & 0.729 & 0.54 \\
0.15 & 146.2 & 124.88 & 114.10 & 1.1 & 1.05 \\
0.10 & 311.62 & 233.2 & 190.41 & 1.7 & 1.64 \\
0.05 & 1244.13 & 913.19 & 663.64 & 5.49 & 6.47 \\
\hline
\end{tabular}
\end{table}
\begin{table}[h!]
\centering
\caption{Computational times (in seconds) for different relative accuracy levels in the Heston digital up-and-in barrier call experiment.}
\label{tab:heston_digital_put}
\begin{tabular}{lrrrrr}
\hline
\textbf{Relative accuracy} & \textbf{MC} & \textbf{MCA} & \textbf{MLMC} & \textbf{AMS 1} & \textbf{AMS 2} \\
\hline
0.20 & 83.24 & 75.94 & 54.12 & 0.64 & 0.65 \\
0.15 & 148.98 & 119.32 & 116.35 & 1.34 & 1.62 \\
0.10 & 322.72 & 238.31 & 196.04 & 4.25 & 3.23 \\
0.05 & 1311.89 & 962.51 & 699.78 & 11.36 & 13.38 \\
\hline
\end{tabular}
\end{table}
\begin{table}[h!]
\centering
\caption{Computational times (in seconds) for different relative accuracy levels in the Black-Scholes Asian digital call experiment.}
\label{tab:bs_Asian_digital_call}
\begin{tabular}{lrrrrrr}
\hline
\textbf{Relative accuracy} & \textbf{MC} & \textbf{MCA} & \textbf{MCV} & \textbf{MLMC} & \textbf{AMS 1} & \textbf{AMS 2} \\
\hline
0.20 & 27.53 & 26.23 &21.77& 15.38 & 1.16 & 0.2 \\
0.15 & 49.85 & 46.84 &40.11& 29.62 & 2.35 & 0.44 \\
0.10 & 105.52 & 92.62 &94.62& 39.69 & 3.85 & 0.79 \\
0.05 & 415.65 & 305.63 &309.81& 228.73 & 9.8 & 15.27 \\
\hline
\end{tabular}
\end{table}

\begin{table}[h!]
\centering
\caption{Computational times (in seconds) for different relative accuracy levels in the Heston Asian digital call experiment.}
\label{tab:heston_Asian_digital_call}
\begin{tabular}{lrrrrrr}
\hline
\textbf{Relative accuracy} & \textbf{MC} & \textbf{MCA} & \textbf{MCV} & \textbf{MLMC} & \textbf{AMS 1} & \textbf{AMS 2} \\
\hline
0.20 & 82.22  & 73.38  &34.31& 53.6  & 4.3  & 2.62    \\
0.15 & 139.31 & 117.6 &60.86& 108.37 & 10.94 & 6.76   \\
0.10 & 301.53 & 228.35 &116.19& 185.88 & 14.08 & 10.79 \\
0.05 & 1221.18& 899.04 &474.85& 640.85 & 106.05  & 36.56  \\
\hline
\end{tabular}
\end{table}
\newpage
\bibliographystyle{plain}
\bibliography{biblio}
\end{document}